\documentclass{article}

\usepackage[T1]{fontenc}
\usepackage[utf8]{inputenc}

\usepackage{ismir}

\usepackage{amsmath,cite,url}
\usepackage{graphicx}
\usepackage{color}
\usepackage{amssymb}
\usepackage{booktabs}
\usepackage{algorithm}
\usepackage{algorithmic}
\usepackage{microtype}

\title{UOT-IR: Structured Routing of High-Polyphony Symbolic Music into Fixed-Budget Representations}

\multauthor
{Ziyue Kang$^{1,2}$ \hspace{0.35cm}
Nan Nan$^{1,2,3*}$ \hspace{0.35cm}
Chenhao Lin$^{1,2,3}$ \hspace{0.35cm}
Xiaohong Guan$^{1,2,3,4}$}
{
$^1$ Frontier Institute of Science and Technology,\\
Xi'an Jiaotong University, Xi'an, China\\
$^2$ Interdisciplinary Research Center of Frontier Science and Technology,\\
Xi'an Jiaotong University, Xi'an, China\\
$^3$ MOE KLINNS Lab, Faculty of Electronic and Information Engineering,\\
Xi'an Jiaotong University, Xi'an, China\\
$^4$ Center for Intelligent and Networked Systems, Tsinghua University, Beijing, China\\
{\tt\small edu.kangziyue@gmail.com, nan.nan@xjtu.edu.cn}
}

\def\authorname{Z. Kang, N. Nan, C. Lin, and X. Guan}

\begin{document}

\maketitle

\begingroup
\renewcommand{\thefootnote}{\fnsymbol{footnote}}
\footnotetext[1]{Corresponding author}
\endgroup

\begin{abstract}
\begingroup
\hyphenpenalty=0
\exhyphenpenalty=0
\looseness=-1
High-polyphony symbolic music is increasingly used in generation, analysis, and arrangement, yet many downstream tasks require bounded representations with fixed tracks or slots. Converting richly orchestrated scores into compact forms is therefore necessary, but existing approaches relying on heuristic simplification or generic representation-space reduction often fail to preserve structural roles, orchestration compatibility, and playability under strict budgets. To address the issue, this study reformulates the compression problem as a fixed-budget structured routing problem and proposes Unbalanced Optimal Transport for Information Routing (UOT-IR), a training-free framework based on constrained unbalanced optimal transport. UOT-IR combines an orchestration prior, adaptive marginal relaxation, temporal decoding, and playability-aware projection to produce compact and musically coherent bounded representations. This work further studies two practical settings under the same slot budget: template standardization, which maps each input to a predefined bounded template, and adaptive preservation, which retains representative content without assuming an external template. Experiments on the SymphonyNet corpus show that UOT-IR delivers strong overall performance across both settings, including the best Note-F1 in adaptive preservation (0.9120), together with the lowest structural cost (14.7165) and bad structural confusion rate (0.3406) in template standardization. This work establishes a principled paradigm for fixed-budget symbolic music compression, offering a practical path toward compact, structured, and musically coherent symbolic representations.
\par
\endgroup
\end{abstract}

\begin{figure*}[t]
    \centering
    \includegraphics[
        width=\textwidth,
        trim=0.1cm 0.1cm 0.1cm 0.2cm,
        clip
    ]{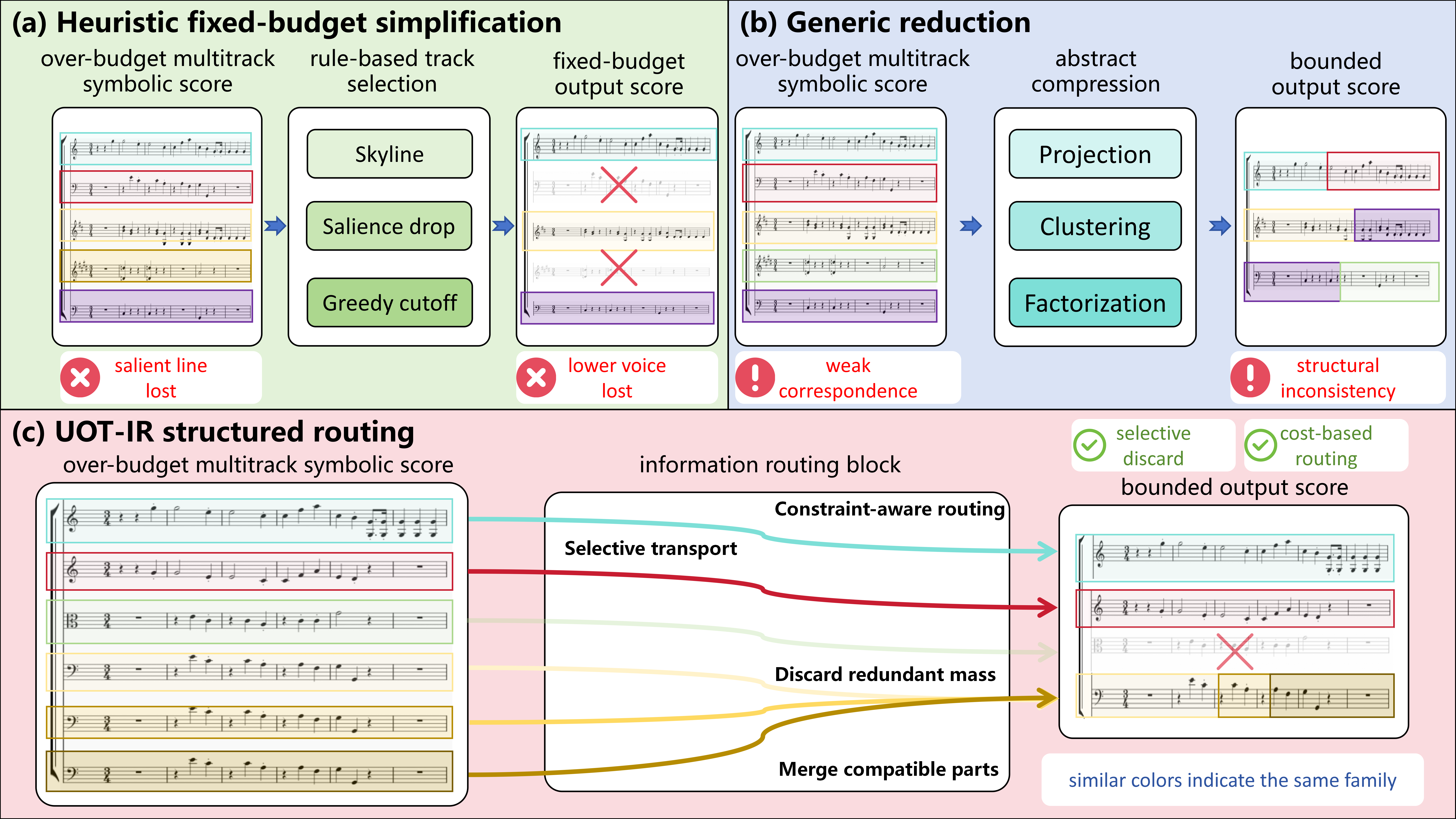}
    \caption{Comparison between conventional fixed-budget compression and UOT-IR. Existing heuristic or representation-space reduction methods often drop salient lines, merge incompatible materials, or violate slot and playability constraints. UOT-IR instead performs structure-aware routing with compatibility-aware assignment, selective discard, and constrained projection to produce coherent bounded outputs.}
    \label{fig:motivation}
\end{figure*}

\section{Introduction}\label{sec:introduction}
High-polyphony symbolic music is increasingly used in generation, analysis, arrangement, and archival settings. In practice, however, many symbolic music representations and processing frameworks impose fixed budgets on the number of tracks, parts, or target slots \cite{Zeng2021MusicBERT,Liu2022SymphonyNet,Ou2024UnifiedArrangement,Wang2025NotaGen}. Once a richly orchestrated score exceeds such a structural budget, it must first be converted into a bounded form.

This conversion problem is nontrivial because source tracks may play different musical roles, carry unequal salience, and remain subject to orchestration and playability constraints. Under a fixed budget, some material should be preserved, some should be reassigned to compatible target roles, and some should be discarded. Existing approaches mainly fall into two categories: heuristic simplification and representation-space reduction. Heuristic strategies such as pruning, merging, or rule-based selection can reduce complexity, but they often treat compression as removing less important material, making it difficult to preserve multiple concurrent functions such as melody, accompaniment, and bass in a coordinated way \cite{Chiu2009PianoReduction,Nakamura2015PianoReduction}. Representation-space reduction methods can also simplify multitrack content, yet they are not explicitly designed to enforce structured source-to-target assignment or playability constraints such as valid pitch range, polyphonic limits, and performance difficulty \cite{Pearson1901PCA,Lee1999NMF,MacQueen1967KMeans,Nakamura2018Difficulty}. Related work on automatic instrumentation further shows that multitrack symbolic music often contains overlapping pitch ranges and dense textures, under which simple assignment strategies can produce unstable or less reliable part allocations \cite{Dong2021Instrumentation}. As illustrated in Fig.~\ref{fig:motivation}, these limitations lead to three practical failure modes: salient lines may be removed, incompatible materials may be merged into the same bounded slot, and the resulting output may violate basic playability or role-consistency requirements. In particular, NotaGen excludes scores with more than 16 staves because of generation complexity \cite{Wang2025NotaGen}.
\looseness=1
To address these limitations, we reformulate fixed-budget symbolic compression as a structured routing problem. UOT-IR is a training-free framework based on constrained unbalanced optimal transport that integrates a taxonomy-grounded orchestration prior, symbolic statistical descriptors, adaptive marginal relaxation, temporal coherence, and playability-aware projection to produce compact yet musically coherent bounded outputs. We study two settings under the same slot budget: template standardization, which maps each input into a predefined bounded template, and adaptive preservation, which preserves representative content without assuming an external template. Experiments on SymphonyNet \cite{Liu2022SymphonyNet} show that UOT-IR achieves favorable overall results against heuristic, representation-space, and simplified transport baselines across both settings.
\looseness=-1
Our contributions are threefold: (1) we formulate fixed-budget symbolic compression as a structured routing problem; (2) we develop a training-free framework that integrates orchestration prior, constrained UOT, adaptive relaxation, temporal decoding, and playability-aware projection; and (3) we evaluate the framework in both settings and show a strong overall balance across fidelity, structural compatibility, and conflict-related metrics.

\section{Related Work}

\subsection{Bounded Symbolic Music Representations}
Many symbolic music generation, arrangement, and analysis frameworks rely on bounded structural representations, such as fixed numbers of tracks, parts, or target slots, to simplify storage, batching, alignment, and model design \cite{Zeng2021MusicBERT,Liu2022SymphonyNet,Ou2024UnifiedArrangement,Wang2025NotaGen,Ji2023SymbolicSurvey}. While practical, such representations create a mismatch between richly orchestrated source scores and normalized bounded formats. Existing work typically assumes that the bounded target format is already given, and therefore focuses on modeling, generation, or rearrangement within that format. By contrast, our focus is the conversion problem that arises before those stages: how to transform over-budget multitrack symbolic music into a bounded representation when the source score exceeds the allowed structural budget.

\subsection{Heuristic and Representation-Space Reduction}
A common way to handle over-budget multitrack music is heuristic reduction, including pruning, merging, or selecting tracks according to activity, density, salience, instrument grouping, or incremental pitch-space coverage \cite{Dong2021Instrumentation,Zhao2023QARearrangement,Ou2024UnifiedArrangement}. Another line performs representation-space approximation through projection, factorization, or clustering, including principal component analysis (PCA), nonnegative matrix factorization (NMF), and K-means \cite{Pearson1901PCA,Lee1999NMF,MacQueen1967KMeans}. These approaches reduce complexity, but they usually do not explicitly model source-to-slot correspondence, orchestration compatibility, and selective discard under a fixed symbolic budget.

\subsection{Optimal Transport for Cost-Aware Partial Matching}
Optimal transport (OT) studies how to align two distributions under an explicit transportation cost, and has become a general framework for cost-aware soft assignment between structured objects \cite{Peyre2019ComputationalOT,Cuturi2013Sinkhorn}. When strict mass conservation is too restrictive, unbalanced optimal transport relaxes the marginal constraints and allows only part of the source mass to be preserved or reassigned \cite{Chizat2018UOT,Chizat2018Scaling}. These properties make UOT a natural tool for symbolic compression under constrained structural budgets, where correspondence, reassignment, and selective discard must all be modeled explicitly.

\section{Method}
\label{sec:method}

\begin{figure*}[!t]
    \centering
    \includegraphics[
        width=\textwidth,
        trim=0.3cm 2.5cm 0.3cm 0.2cm,
        clip
    ]{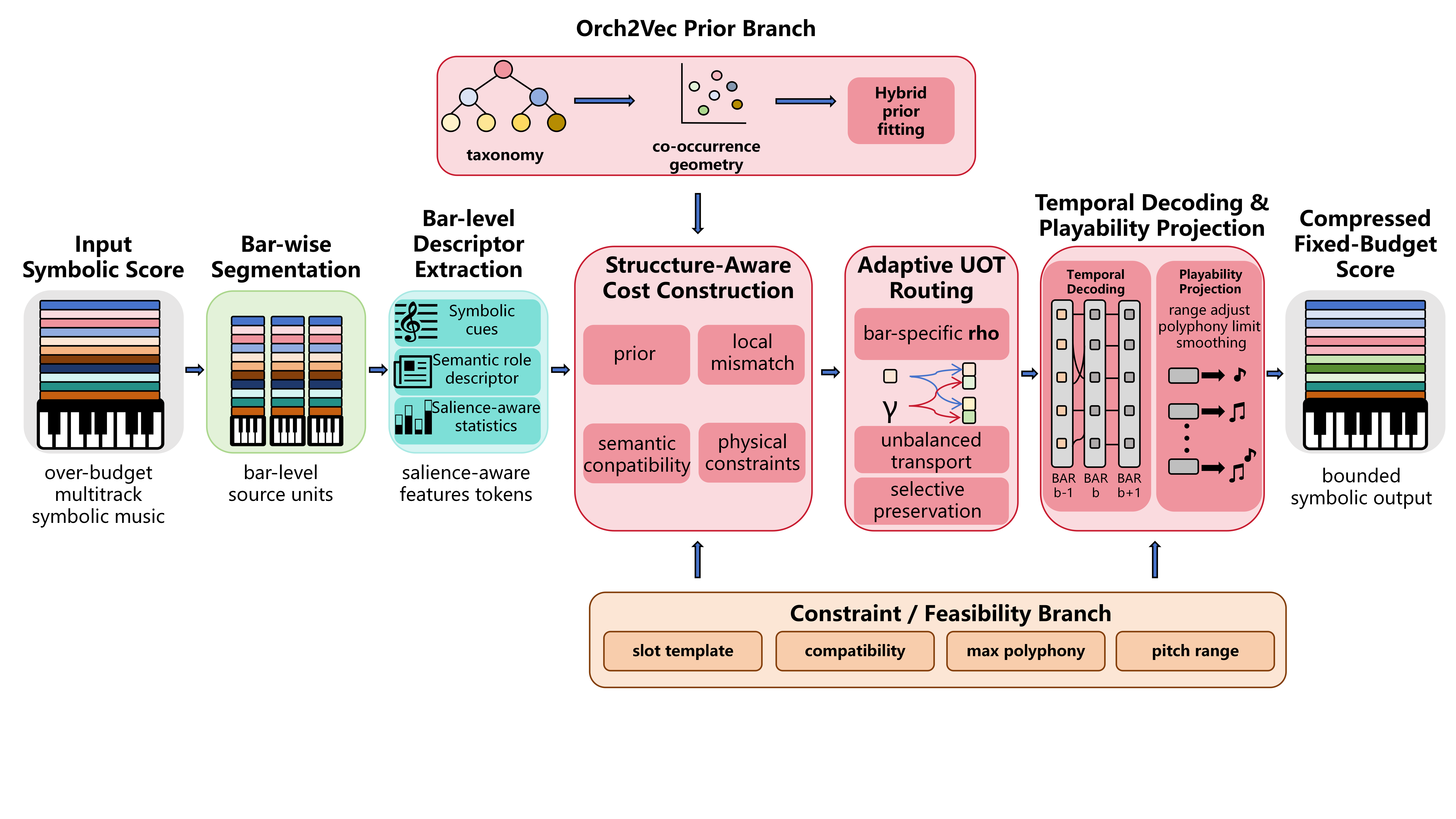}
    \caption{Overview of UOT-IR. The framework consists of five conceptual stages: problem formulation, descriptor and marginal construction, structure-aware cost construction, adaptive UOT routing, and temporal decoding with playability-aware projection. These stages are described across the subsections of Section~\ref{sec:method}.}
    \label{fig:framework}
\end{figure*}

UOT-IR instantiates the structured-routing view of fixed-budget symbolic compression. Given an over-budget multitrack symbolic score, the goal is to construct a bounded symbolic output that preserves musically salient content, maintains structural consistency, and reduces physically implausible assignments, as illustrated in Fig.~\ref{fig:motivation}. Instead of treating this task as heuristic track reduction or generic representation-space approximation, the method models it as a structure-aware routing problem under constrained unbalanced optimal transport (UOT). An overview of the full pipeline is shown in Fig.~\ref{fig:framework}.

\subsection{Problem Formulation}
We segment a multitrack symbolic piece into bars indexed by $b \in \{1,\dots,B\}$, where $B$ is the total number of bars. For bar $b$, let $X^{(b)}=\{x^{(b)}_1,\dots,x^{(b)}_{N_b}\}$ denote the set of active source tracks and $Y^{(b)}=\{y^{(b)}_1,\dots,y^{(b)}_K\}$ denote the set of $K$ target slots under the fixed compression budget, where $K$ is constant across all bars. In \emph{template standardization}, these slots are predefined by a canonical bounded template. In \emph{adaptive preservation}, they are instantiated from representative source-derived content under the same budget.

For each bar, UOT-IR solves for a nonnegative transport matrix $\Gamma^{(b)} \in \mathbb{R}_+^{N_b \times K}$, where $\Gamma^{(b)}_{ik} \ge 0$ denotes the transported mass from source track $x_i^{(b)}$ to target slot $y_k^{(b)}$. Quantitatively,
\begin{equation}
\begin{aligned}
s_i^{(b)}
&= \sum_{k=1}^{K}\Gamma_{ik}^{(b)}, \quad
c_k^{(b)}
= \sum_{i=1}^{N_b}\Gamma_{ik}^{(b)}, \\
m_b
&= \|\Gamma^{(b)}\|_1
= \sum_{i=1}^{N_b}\sum_{k=1}^{K}\Gamma_{ik}^{(b)}.
\end{aligned}
\label{eq:mass_definitions}
\end{equation}
Here, $s_i^{(b)}$ is the retained mass of source track $x_i^{(b)}$, $c_k^{(b)}$ is the received mass of target slot $y_k^{(b)}$, and $m_b$ is the total retained mass at bar $b$.

\subsection{Descriptor and Marginal Construction}
Each source track $x_i^{(b)}$ is represented by a bar-level descriptor, while each target slot is associated with a piece-level prototype descriptor
\begin{equation}
\phi_i^{(b)} \in \mathbb{R}^{d},
\qquad
\psi_k \in \mathbb{R}^{d},
\label{eq:descriptors}
\end{equation}
respectively, where $d$ is the shared descriptor dimension. The source descriptors summarize bar-level symbolic statistics, including pitch, duration, onset, density, polyphony, and rhythmic characteristics. To determine how much material should be routed from each source track, we define a nonnegative source marginal
\begin{equation}
\mu^{(b)} \in \mathbb{R}_+^{N_b}.
\label{eq:source_marginal}
\end{equation}
Here, $\mu_i^{(b)}$ is the source mass of track $x_i^{(b)}$, and $\nu_k^{(b)}$ is the receiving mass of target slot $y_k^{(b)}$. In implementation, $\mu^{(b)}$ is computed from bar-level activity and salience cues and then normalized, while the target marginal is uniform, i.e., $\nu_k^{(b)}=1/K$.

A key component of UOT-IR is \textbf{Orch2Vec}, a program-level orchestration prior. We define a vocabulary of 129 instrument tokens, consisting of 128 General MIDI programs and one dedicated drums token. These tokens are organized in a tree-structured taxonomy spanning root-level families (e.g., Strings, Winds, Keys\_Plucked, Percussion\_Family, and Synth\_SFX), intermediate subfamilies, and leaf-level program identities. To complement this symbolic structure, we derive a data-driven program geometry from corpus-level co-occurrence statistics. Let $a$ and $b$ denote two instrument programs, $P(a,b)$ their joint co-occurrence probability, and $P(a)$ and $P(b)$ their marginal probabilities. We compute the positive pointwise mutual information (PPMI) \cite{ChurchHanks1990PMI} as
\begin{equation}
\mathrm{PPMI}(a,b)
=
\max\left\{
\log \frac{P(a,b)}{P(a)P(b)},\, 0
\right\}.
\label{eq:ppmi}
\end{equation}
In implementation, the empirical co-occurrence geometry is further stabilized by support gating and clipping before pairwise distances are computed. We then combine this empirical geometry with the taxonomy graph by fitting tree-consistent edge weights and deriving a symmetric prior matrix
\begin{equation}
C_{\text{prior}} \in \mathbb{R}^{129 \times 129},
\label{eq:prior_matrix}
\end{equation}
which assigns lower routing cost to source--target program pairs that are both statistically compatible and structurally close in the taxonomy.

\subsection{Structure-Aware Cost Construction and UOT Routing}
This prior matrix serves as the orchestration-aware component in the subsequent routing-cost decomposition, where it is combined with local, semantic, and physical terms. Given the routing cost matrix
\begin{equation}
C^{(b)} \in \mathbb{R}^{N_b \times K},
\label{eq:cost_matrix}
\end{equation}
we solve for a nonnegative transport matrix by
\begin{equation}
\begin{aligned}
\min_{\Gamma^{(b)} \ge 0}\quad
& \left\langle \Gamma^{(b)}, C^{(b)} \right\rangle
+ \lambda_s D_{\rho_b}\!\left(
\Gamma^{(b)}\mathbf{1}, \mu^{(b)}
\right) \\
& + \lambda_t D_{\rho_b}\!\left(
(\Gamma^{(b)})^\top\mathbf{1}, \nu^{(b)}
\right),
\end{aligned}
\label{eq:uot_main}
\end{equation}
where $\langle\Gamma^{(b)},C^{(b)}\rangle$ is the total routing cost, $\mathbf{1}$ is an all-ones vector, and $\lambda_s$ and $\lambda_t$ weight deviations of the transported source and target marginals, $\Gamma^{(b)}\mathbf{1}$ and $(\Gamma^{(b)})^\top\mathbf{1}$, from $\mu^{(b)}$ and $\nu^{(b)}$, respectively. The parameter $\rho_b$ controls marginal relaxation, allowing selective discard under the fixed target budget. A standard choice for the unbalanced penalty is the generalized Kullback--Leibler divergence
\begin{equation}
D_{\rho_b}(\mathbf{a},\mathbf{b})
=
\rho_b \sum_j
\left(
a_j \log \frac{a_j}{b_j}
- a_j + b_j
\right),
\label{eq:kl_unbalanced}
\end{equation}
where $\mathbf{a}$ and $\mathbf{b}$ are nonnegative vectors of equal dimension, $a_j$ and $b_j$ are their $j$-th entries, and $\rho_b>0$ controls the strength of marginal relaxation.

The routing cost matrix is conceptually decomposed as
\begin{equation}
C^{(b)}
=
\alpha C_{\text{prior}}^{(b)}
+
\beta C_{\text{local}}^{(b)}
+
\gamma C_{\text{semantic}}^{(b)}
+
\delta C_{\text{physical}}^{(b)}.
\label{eq:cost_decomposition}
\end{equation}
Here, the four components capture program compatibility, descriptor mismatch, role mismatch, and physical infeasibility, respectively, while the conceptual weights $\alpha,\beta,\gamma,\delta\geq0$ control their relative contributions. In practice, these components are fused through a semantic--local fusion weight, prior--semantic gating, and penalty scaling, which are partially refined by lightweight test-time adaptation (TTA).

The prior term is
\begin{equation}
(C_{\text{prior}}^{(b)})_{ik}=C_{\text{prior}}[p_i,q_k],
\label{eq:prior_term}
\end{equation}
where $p_i$ and $q_k$ denote the MIDI programs of source track $x_i^{(b)}$ and target slot $y_k^{(b)}$, respectively. The local term is
\begin{equation}
(C_{\text{local}}^{(b)})_{ik}
=
d_{\text{feat}}\!\left(\phi_i^{(b)},\psi_k\right),
\label{eq:local_term}
\end{equation}
where $d_{\text{feat}}$ is implemented as cosine distance. Thus, Equation~\eqref{eq:local_term} measures the distance between the descriptor $\phi_i^{(b)}$ of source track $x_i^{(b)}$ and the prototype descriptor $\psi_k$ of target slot $y_k$. In the current implementation, target prototypes are initialized from piece-level source-track descriptors and then used as the slot-side references for bar-level routing.

The semantic term is
\begin{equation}
(C_{\text{semantic}}^{(b)})_{ik}
=
d_{\text{sem}}(r_i^{(b)},u_k^{(b)}),
\label{eq:semantic_term}
\end{equation}
where $r_i^{(b)}$ and $u_k^{(b)}$ denote semantic embeddings derived from the source-track and target-slot names, respectively, thereby providing coarse role-related cues for routing. The physical term is
\begin{equation}
\begin{aligned}
(C_{\text{physical}}^{(b)})_{ik}
=
\eta_{\text{range}}
\Pi_{\text{range}}(x_i^{(b)},y_k^{(b)}),
\end{aligned}
\label{eq:c_physical}
\end{equation}
where the routing-stage physical term captures pitch-range compatibility between the source material and the target slot.
\begin{equation}
\begin{aligned}
\Pi_{\text{range}}(x_i^{(b)},y_k^{(b)})
={}&
\max(0,\, \ell_k - p_{i,\min}^{(b)}) \\
&+ \max(0,\, p_{i,\max}^{(b)} - u_k),
\end{aligned}
\label{eq:range_penalty}
\end{equation}
where $[p_{i,\min}^{(b)}, p_{i,\max}^{(b)}]$ is the observed pitch interval of the source track and $[\ell_k,u_k]$ is the admissible pitch range of the target slot.

\subsection{Adaptive Relaxation, Temporal Decoding, and Playability-Aware Projection}
Different bars may require different retention strength, so using a single fixed unbalancedness parameter for all bars is suboptimal. UOT-IR therefore uses a bar-specific relaxation parameter $\rho_b$. We choose $\rho_b$ by matching the retained mass to a desired target level:
\begin{equation}
\rho_b^\star
=
\arg\min_{\rho \in \mathcal{R}}
\left|
\|\Gamma^{(b)}(\rho)\|_1 - \tau_b
\right|,
\label{eq:rho_search}
\end{equation}
where $\rho_b^\star$ is the selected relaxation strength, $\mathcal{R}$ is the search range, $\Gamma^{(b)}(\rho)$ is the transport plan obtained with candidate $\rho$, and $\tau_b$ is the desired retained mass for bar $b$. In implementation, the current solver performs a lightweight secant-style search in log-$\rho$ space on top of the refined cost matrix, so that each bar obtains its own relaxation strength before the final UOT solve.

Because the transport problem in Equation~\eqref{eq:uot_main} is solved independently for each bar, the resulting slot identities may fluctuate over time. We therefore convert the bar-wise routing solution into temporally coherent slot sequences through a target-aware sticky decoding scheme, implemented as a Viterbi-style dynamic program \cite{Viterbi1967}:
\begin{equation}
\max_{\{z_i^{(b)}\}}\;
\sum_{b,i} E_b(i,z_i^{(b)})
-\sum_{b>1,i}T\!\left(z_i^{(b-1)},z_i^{(b)}\right),
\label{eq:viterbi}
\end{equation}
where $z_i^{(b)}$ denotes the target slot assigned to source track $i$ at bar $b$, and $E_b(i,\cdot)$ is the corresponding emission score derived from the routing result.
We define the transition cost between target slots $a$ and $b$ as
\begin{equation}
\begin{aligned}
T(a,b)
={}&
\lambda_{\text{stay}} \, \mathbf{1}[a \neq b] \\
&+ \lambda_{\text{prog}} \,
\mathbf{1}[\mathrm{prog}(a) \neq \mathrm{prog}(b)].
\end{aligned}
\label{eq:transition}
\end{equation}
It penalizes unnecessary slot switching and target-program changes. During decoding, its strength is further modulated by track importance derived from the accumulated source marginal, improving stability for musically salient tracks.

After temporal decoding, routed note events are projected into target slots with playability-aware refinement, including pitch-range correction, short-note filtering, and target-dependent polyphony control. The final compressed symbolic output for bar $b$ is
\begin{equation}
\hat{Y}^{(b)}
=
\{\hat{\mathcal{N}}_1^{(b)}, \dots, \hat{\mathcal{N}}_K^{(b)}\},
\label{eq:final_output}
\end{equation}
where $\hat{\mathcal{N}}_k^{(b)}$ denotes the refined note set assigned to target slot $k$ in bar $b$. Algorithm~\ref{alg:UOT-IR} summarizes the full UOT-IR pipeline.

\begin{algorithm}[H]
\caption{UOT-IR for fixed-budget symbolic compression}
\label{alg:UOT-IR}
\footnotesize
\begin{algorithmic}[1]
\REQUIRE Multitrack symbolic piece $\mathcal{X}$, slot budget $K$, target regime, orchestration prior $C_{\text{prior}}$
\ENSURE Bounded symbolic output $\hat{\mathcal{Y}}$
\STATE Segment $\mathcal{X}$ into bars
\STATE Instantiate target slots according to the regime
\FOR{each bar $b$}
    \STATE Extract active source tracks and bar-level descriptors
    \STATE Compute source and target marginals
    \STATE Build and refine the routing cost matrix
    \STATE Search bar-specific relaxation parameter $\rho_b$
    \STATE Solve constrained UOT to obtain $\Gamma^{(b)}$
\ENDFOR
\STATE Apply temporal decoding across bars
\STATE Project routed note events with playability-aware refinement
\STATE Return $\hat{\mathcal{Y}}$
\end{algorithmic}
\end{algorithm}

\section{Experimental Setup}

\subsection{Dataset and Evaluation Protocol}
We evaluate UOT-IR on the SymphonyNet corpus under a fixed slot budget. From the full symbolic collection, we construct evaluation subsets by selecting pieces or bars whose active track count exceeds the target budget, so that compression is required rather than trivial identity mapping. We consider two settings under the same budget $K$. In \textbf{template standardization}, each input is mapped to a predefined bounded target template. In \textbf{adaptive preservation}, each method compresses the same input under the same budget without assuming an external template. Unless otherwise stated, all compared methods operate on the same segmented bar-level inputs to ensure fair comparison across settings.

\subsection{Baselines}
We compare UOT-IR against three groups of baselines. \textbf{Heuristic baselines} include \emph{Direct}, \emph{Random}, \emph{Greedy}, \emph{Skyline}, and \emph{BMF-PC}. These are task-level reference strategies constructed for this study, covering direct retention, random selection, salience-driven selection, rule-based reduction, and incremental pitch-space coverage. Specifically, Skyline prioritizes the drum track and the source tracks at the lower and upper register extremes, and then fills the remaining budget according to note density. Boolean matrix pitch-coverage (BMF-PC) is a coverage-based heuristic that iteratively selects representative tracks according to their incremental pitch-space contribution. \textbf{Representation-space baselines} include \emph{PCA} \cite{Pearson1901PCA}, \emph{NMF} \cite{Lee1999NMF}, and \emph{KMeans} \cite{MacQueen1967KMeans}, which select representative source tracks through projection, factorization, or clustering in a shared structured feature space. These methods are then converted into bounded symbolic outputs for evaluation. \textbf{Transport baselines} include \emph{Vanilla-UOT} \cite{Chizat2018UOT,Chizat2018Scaling}, which removes the structure-aware components from our framework, and \emph{UOT-IR-Core}, which retains the transport backbone but omits selected refinement modules.

In template standardization, methods that do not natively produce slot-aligned outputs are post-aligned to the target template using Hungarian matching \cite{Kuhn1955Hungarian}. In adaptive preservation, no such post-alignment is applied.

\subsection{Metrics}
We evaluate all methods from two perspectives. \textbf{Content fidelity} is measured by note-level precision, recall, and F1 (Note-P, Note-R, and Note-F1), together with Fréchet Track Embedding Distance (FTED) and Jensen--Shannon distance (JSD) over pitch-class, duration, and inter-onset-interval distributions (PC-JSD, Dur-JSD, and IOI-JSD). \textbf{Structural compatibility} is measured by structural cost (SC), bad structural confusion rate (BC), pitch-range difference (PR-Diff), and pitch-class-entropy difference (PCE-Diff). SC measures the average Orch2Vec program cost between matched source and output notes, while BC measures the rate of matched notes whose Orch2Vec program cost exceeds the structural-confusion threshold. Lower FTED, JSD, SC, BC, PR-Diff, and PCE-Diff values indicate better preservation or structural compatibility.
\section{Results and Analysis}

\subsection{Results on Adaptive Preservation}

Table~\ref{tab:main_adaptive} reports results on adaptive preservation. UOT-IR achieves the best overall Note-F1 and recall, indicating a stronger balance between content retention and bounded-output coherence. Heuristic baselines often obtain very high precision by keeping only low-risk content, but their lower recall limits overall usefulness. Compared with representation-space baselines, the gains suggest that fixed-budget symbolic compression requires correspondence-aware routing rather than generic low-dimensional approximation alone.
\begin{table}[!htbp]
    \centering
    \caption{Main results on the adaptive reduction task.}
    \label{tab:main_adaptive}
    \small
    \setlength{\tabcolsep}{3.0pt}
    \resizebox{\columnwidth}{!}{%
    \begin{tabular}{lccc|cccc}
        \toprule
        & \multicolumn{3}{c|}{\textbf{Content Fidelity}} & \multicolumn{4}{c}{Distributional and Structural Statistics} \\
        \cmidrule(lr){2-4}\cmidrule(lr){5-8}
        Method & Note-F1$^{*}$ $\uparrow$ & Note-P $\uparrow$ & Note-R$^{\dagger}$ $\uparrow$ & FTED $\downarrow$ & PC-JSD $\downarrow$ & Dur-JSD $\downarrow$ & IOI-JSD $\downarrow$ \\
        \midrule
        Direct      & 0.7709 & 0.9999 & 0.6563 & 0.0184 & 0.1049 & 0.1282 & 0.1328 \\
        Random      & 0.7757 & 0.9999 & 0.6545 & 0.0099 & 0.0768 & 0.0790 & 0.1094 \\
        Greedy      & 0.9113 & 0.9998 & 0.8458 & 0.0207 & 0.0376 & 0.0562 & 0.0411 \\
        Skyline     & 0.8778 & 0.9975 & 0.7982 & 0.0138 & 0.0369 & 0.0618 & 0.0481 \\
        PCA         & 0.7092 & 0.9998 & 0.5813 & 0.0071 & 0.0787 & 0.0918 & 0.1244 \\
        KMeans      & 0.7338 & 0.9999 & 0.6046 & \textbf{0.0054} & 0.0645 & 0.0711 & 0.1139 \\
        BMF-PC      & 0.8989 & 0.9998 & 0.8270 & 0.0172 & 0.0356 & 0.0554 & 0.0444 \\
        NMF         & 0.7401 & 0.9999 & 0.6152 & 0.0060 & 0.0672 & 0.0782 & 0.1102 \\
        \textbf{Vanilla-UOT} & 0.8593 & 0.8896 & 0.8419 & 0.0351 & 0.0109 & 0.0234 & 0.0130 \\
        \textbf{UOT-IR-Core} & 0.9053 & 0.9104 & 0.9082 & 0.0228 & \textbf{0.0041} & \textbf{0.0142} & \textbf{0.0056} \\
        \textbf{UOT-IR}      & \textbf{0.9120} & 0.9129 & \textbf{0.9200} & 0.0230 & 0.0071 & 0.0178 & 0.0078 \\
        \bottomrule
    \end{tabular}%
    }
\end{table}

Several observations are worth highlighting. First, some baselines achieve nearly perfect precision because they retain only highly confident or low-risk content, but this comes at the cost of much lower recall and therefore lower overall usefulness under fixed-budget preservation. Second, the transport-based variants achieve low values on several distributional metrics, showing that transport-based routing effectively preserves coarse bar-level statistics. However, their weaker overall balance indicates that matching distributions is not sufficient: compression also requires musically structured routing and robust content retention. Third, the full UOT-IR model achieves the best overall Note-F1 and recall, showing that the complete routing formulation provides a strong trade-off between preserving important material and maintaining coherent bounded outputs.

\subsection{Main Results on Template Standardization}

Table~\ref{tab:main_standardization} reports results on template standardization. UOT-IR achieves the best overall task-relevant performance, especially on PR-Diff, PCE-Diff, SC, and BC. Although it does not always obtain the lowest FTED, this is expected because template standardization requires not only structural similarity but also template compatibility and low conflict. The strong gap over Vanilla-UOT further shows that transport alone is insufficient under strict template constraints.

\begin{table}[!htbp]
    \centering
    \caption{Main results on the standardization task.}
    \label{tab:main_standardization}
    \small
    \setlength{\tabcolsep}{3.0pt}
    \resizebox{\columnwidth}{!}{%
    \begin{tabular}{l c|ccccc}
        \toprule
        & \multicolumn{1}{c|}{Fidelity} & \multicolumn{5}{c}{\textbf{Standardization and Structural Compatibility}} \\
        \cmidrule(lr){2-2}\cmidrule(lr){3-7}
        Method & Note-F1 $\uparrow$ & PR-Diff$^{*}$ $\downarrow$ & PCE-Diff$^{*}$ $\downarrow$ & FTED $\downarrow$ & SC$^{*}$ $\downarrow$ & BC$^{*}$ $\downarrow$ \\
        \midrule
        Direct      & 0.6712 & 0.1421 & 0.0699 & 0.0186 & 20.3849 & 0.4681 \\
        Random      & 0.6732 & 0.0953 & 0.0824 & 0.0100 & 18.1475 & 0.4247 \\
        Greedy      & 0.7585 & 0.0514 & 0.0745 & 0.0209 & 19.1898 & 0.4645 \\
        Skyline     & 0.7301 & 0.0568 & 0.0864 & 0.0140 & 18.2123 & 0.4451 \\
        PCA         & 0.6078 & 0.1218 & 0.1210 & 0.0073 & 16.6987 & 0.4116 \\
        KMeans      & 0.6327 & 0.1009 & 0.1041 & \textbf{0.0056} & 16.5824 & 0.3939 \\
        BMF-PC      & 0.7515 & 0.0501 & 0.0799 & 0.0173 & 18.8827 & 0.4570 \\
        NMF         & 0.6334 & 0.1099 & 0.1077 & 0.0062 & 16.8215 & 0.4077 \\
        \textbf{Vanilla-UOT} & 0.6335 & 0.0461 & 0.0842 & 0.0294 & 30.2076 & 0.7695 \\
        \textbf{UOT-IR-Core} & 0.9334 & 0.0196 & 0.0400 & 0.0225 & 18.1452 & 0.4466 \\
        \textbf{UOT-IR}      & \textbf{0.9370} & \textbf{0.0116} & \textbf{0.0205} & 0.0238 & \textbf{14.7165} & \textbf{0.3406} \\
        \bottomrule
    \end{tabular}%
    }
\end{table}

\begin{table}[!htbp]
    \centering
    \caption{Ablation results on standardization and adaptive preservation.}
    \label{tab:ablation_all}
    \footnotesize
    \setlength{\tabcolsep}{2.6pt}
    \renewcommand{\arraystretch}{0.95}
    \resizebox{\columnwidth}{!}{%
    \begin{tabular}{lccccccc}
        \toprule
        & \multicolumn{3}{c}{Standardization} & \multicolumn{4}{c}{Adaptive} \\
        \cmidrule(lr){2-4}\cmidrule(lr){5-8}
        Variant & Note-F1$^{*}$ $\uparrow$ & SC$^{*}$ $\downarrow$ & BC$^{*}$ $\downarrow$ & Note-F1$^{*}$ $\uparrow$ & FTED $\downarrow$ & SC$^{*}$ $\downarrow$ & BC$^{*}$ $\downarrow$ \\
        \midrule
        UOT-IR            & \textbf{0.9370} & 14.7165 & 0.3406 & 0.9120 & 0.0230 & \textbf{2.0296} & \textbf{0.0575} \\
        w/o Prior        & 0.8317          & 26.9149 & 0.7528 & \textbf{0.9310} & 0.0162 & 13.4090 & 0.2643 \\
        w/o Symb. Stat.  & 0.8853          & 18.6693 & \textbf{0.3403} & 0.8131 & 0.0257 & 3.8790 & 0.0783 \\
        w/o UOT          & 0.9146          & 18.5173 & 0.4754 & 0.7929 & 0.0237 & 9.1925 & 0.3548 \\
        w/o Ada.\ $\rho$ & 0.8601          & \textbf{13.1580} & 0.3808 & 0.8838 & \textbf{0.0150} & 2.8754 & 0.0614 \\
        w/o Temp.        & 0.9342          & 14.6388 & 0.3492 & 0.8750 & 0.0189 & 2.4669 & 0.0782 \\
        w/o TTA          & 0.9337          & 18.1408 & 0.4462 & 0.7978 & 0.0236 & 6.9246 & 0.2594 \\
        \bottomrule
    \end{tabular}%
    }
\end{table}

A notable result is that UOT-IR does not always achieve the lowest FTED. This is reasonable because FTED captures only one aspect of structural similarity, whereas the standardization task also requires template compatibility, low conflict, and musically plausible slot assignment. In this setting, SC and BC are particularly informative, since they quantify the program-level compatibility of matched source and output notes under the bounded target template. The comparison with Vanilla-UOT is also instructive: without the structure-aware components, a plain transport formulation remains insufficient under a strict template constraint, leading to much worse SC and BC despite using the same transport backbone.

\subsection{Ablation Study}
To verify the contributions of the main components, we conduct ablation studies under both settings. As shown in Table~\ref{tab:ablation_all}, the prior improves structural compatibility, while UOT and TTA are important for adaptive preservation; overall, the full model achieves the best balance across metrics.
\section{Conclusion}
This paper shows that over-budget symbolic music is better modeled as a structured routing problem than as heuristic simplification or generic representation-space reduction. UOT-IR is a training-free framework based on constrained unbalanced optimal transport. It unifies template standardization and adaptive preservation within a single formulation. Experiments on the SymphonyNet corpus show competitive overall results against heuristic, representation-space, and simplified transport baselines in both settings. These results establish a practical routing-based paradigm for converting high-polyphony symbolic music into bounded yet musically meaningful representations. Future work will study bounded routing-based representations for controllable arrangement, orchestration-aware generation, and fixed-budget symbolic music modeling.

\section{ACKNOWLEDGMENTS}

This work was supported in part by the National Natural Science Foundation
of China under Grants 62303373, T2341003, and 62376210, and in part by the
XJTU Research Fund for AI Science under Grant 2025YXYC011.

\section{Ethics Statement}

This work uses a third-party symbolic music dataset in accordance with its applicable licenses and usage conditions. The proposed method produces fixed-budget symbolic music representations for research on music information routing. The interpretation and use of the resulting representations should consider the musical context, dataset characteristics, and target representation settings.
\section{AI Usage Statement}

Generative AI tools were used for language polishing, formatting assistance, and routine coding support.

\bibliography{ISMIRtemplate}

\end{document}